\documentclass[preprint,aps,prd,showpacs,nofootinbib, twocolumn,tightenlines]{revtex4-1}
\usepackage{dcolumn}
\usepackage{natbib}
\usepackage{amsmath}
\usepackage{amssymb}
\usepackage{amsfonts}
\usepackage{CJK}
\usepackage{color}
\usepackage{graphicx}
\usepackage{float}
\usepackage{epsfig}
\usepackage{psfrag}
\usepackage{fullpage}
\usepackage{appendix}
\newcommand{\beq}{\begin{equation}}
\newcommand{\eeq}{\end{equation}}
\newcommand{\ba}{\begin{eqnarray}}
\newcommand{\ea}{\end{eqnarray}}
\newcommand{\beal}{\begin{align}}
\newcommand{\eeal}{\end{align}}
\begin{document}
\title{Suggestion for measuring the weak dipole moment of $\tau$ lepton at Z-factory}
\author{Qing-Jun Xu$^{1,3}$}
\email{xuqingjun@hznu.edu.cn}
 \author{Chao-Hsi Chang$^{2,3}$}
\email{zhangzx@itp.ac.cn}
\affiliation{$^{1}$~Department of Physics, Hangzhou Normal University, Hangzhou
310036, China\\
$^{2}$ CCAST (World Laboratory), P.O.Box 8730, Beijing 100190, China\\
$^{3}$ State Key Laboratory of Theoretical Physics, Institute of Theoretical Physics, Chinese Academy of Sciences, Beijing 100190, China }
\begin{abstract}
Three new observables for Z-factory to measure the weak dipole moment of $\tau$ lepton ($d_w^\tau$) are proposed. As those observables
employed at LEP-I, the new observables depend on the real and imaginary parts of $d_w^\tau$ linearly, thus we calculate the slopes, i.e. the linear coefficients of the dependence, precisely and find that the obtained slopes are much lager than those employed at LEP-I. It means that the signal for the weak dipole moment $d_w^\tau$ may be enhanced quite a lot, if one employs the new observables. Being superior to those at LEP-I, we recommend measuring the weak dipole moment $d_w^\tau$ in terms of the new observables if re-analyzing the data of LEP-I or doing the measurement at a possible Z-factory in the near future.
\\

\end{abstract}
\maketitle

\section{Introduction}
The weak electric dipole moment of a charged lepton $l$ ($l=e\,,\mu\,,\tau$),
$d_w^l$, which is called as the
`weak diploe moment' of the lepton $l$ in this paper,
merges in the $Zl^-l^+$ coupling, hence the most sensitive place to observe it
is at an $e^+e^-$ collider running at center-of-mass energy $\sqrt S = m_Z$
i.e. a Z-factory.
At born level of the Standard Model (SM), the dipole moment is zero, though
high order correction may make it nonzero. In fact,
the prediction of SM is that it is far below the present experiment
sensitivity. Thus a sizable nonzero
signal of $d_w^l$ will be an unambiguous indication for new physics beyond SM.
Moreover, the weak dipole moment $d_{w}^{l}$ is also used to parameterize a kind of
the CP-violating effects relevant to $l$ lepton and $Z$-boson, which is known
in many models beyond SM \citep{Bernreuther:1996dr, Hollik:1997ph}
to be proportional to the mass of the lepton $l$.
Among the charged leptons, the lepton $\tau$ is the heaviest one and `unstable' (in detector size),
thus the observation of $d_w^\tau$ is of specially interesting.
For this reason, LEP-I has made a lot of efforts on the
observation of the weak dipole moment of $\tau$-lepton $d_w^\tau$
\citep{Buskulic:1992jj, Buskulic:1994yh, Acton:1992ff, Akers:1994xf, Ackerstaff:1996gy, Acciarri:1998zc, Heister:2002ik}.
Therefore, it is sensible to improve the observation of the weak dipole moment
$d_w^\tau$ further, for re-analyzing the data of LEP-I
and/or for observing it at a new $Z$-factory with a much high luminosity.
Note that, besides ILC, such a new $Z$-factory is indeed under consideration
now \citep{Ma:2010Zfactory}. Since we believe that the $\tau$ weak dipole moment
$d_w^\tau$ may offer hints on the CP-violation beyond the CKM phase in
comparison with the other leptons, and precisely measuring it may help
to search for clues of new physics beyond SM, we would like
to highlight the observation of the $\tau$ weak dipole moment $d_w^\tau$ in this paper.

The observations of the weak dipole moment $d_w^\tau$ at LEP-I
are based on CP-odd observables \citep{Bernreuther:1989kc,Bernreuther:1991xe,Bernreuther:1993nd}, which are constructed by the unit momenta
of the decay-products of $\tau^+$ and $\tau^-$, and it is found that
the mean values of the observables to be proportional to the real part of the weak dipole moment $d_w^\tau$.
Namely in the most methods adopted by LEP-I, the flying directions of $\tau$ decay-products are used
for constructing the observables,
but the information about the momenta and spins of the $\tau$ leptons is not used at all.
Only two optimal observables, which depend on the directions and spins of the $\tau$ leptons, were
employed in ref.\citep{Akers:1994xf}, where both of the real and imaginary parts of $d_w^\tau$
were measured. The useful investigation for experimental measurements on the differential cross sections with different
spins of $\tau$ leptons, as well as an asymmetry, which depend
on the directions of $\tau$ leptons and their products etc, may be found in refs. \citep{Bernabeu:1993er, Stiegler:1993hi}.
Based on these theoretical consideration, the weak dipole moment $d_w^\tau$ were measured by the ALEPH, OPAL and L3 collaborations
at LEP-I \citep{Buskulic:1992jj, Buskulic:1994yh, Acton:1992ff, Akers:1994xf, Ackerstaff:1996gy, Acciarri:1998zc, Heister:2002ik},
and as results and examples of the experimental observations, the best bound on $d_{w}^{\tau}$
\begin{eqnarray}
{\rm Re}(d_{w}^{\tau}) &<& 0.5\times 10^{-17}\ \ {\rm e\ cm}, \nonumber \\
{\rm Im}(d_{w}^{\tau}) &< &1.1 \times 10^{-17}\ \ {\rm e\ cm}
\label{constraints}
\end{eqnarray}
was reached by the ALPEH collaboration in 2003 \citep{Heister:2002ik}.

In this paper, based on investigating the observables which describe the asymmetries
relating to T-odd and T-even operators and to pursue improving the sensibility
in measuring the weak dipole moment $d_w^\tau$, we will show that at least three kinds
of observables may raise the sensibility.

In order to suppress the background from SM, the relevant observables are
constructed by applying the asymmetries related to particles and their antiparticles etc.
Furthermore, phase space of $\tau$-production processes are divided into two parts for the purpose of
enhancing the signal of these observables. These newly constructed observables, to which
the SM contribution is canceled, can be expressed as the linear functions of
the real and imaginary parts of the weak dipole moment $d_w^\tau$. Their slope coefficients, i.e. the
dependence on the couplings of $Z$ to leptons, the kinematics of the processes, as well as the
decay modes of $\tau$-lepton, are calculated numerically. In comparison with the observables
employed at LEP-I, it is shown that the new observables reflecting the asymmetries are
much more sensitive to the real and imaginary parts of the weak dipole moment $d_{w}^{\tau}$ than
those employed at LEP-I.

The paper is organized as following. In section II three new kinds of observables are constructed,
and important properties of them are pointed out. Observables employed at LEP-I are reviewed in section III.
In section IV, the precise numerical values for the slope coefficients of the new observables,
as well as those employed at LEP-I, are presented, and comparisons with that of previous observables
are made. Brief discussions and conclusion are put at the end of this section.

\section{New observables sensitive to the weak dipole moment $d_{w}^{\tau}$}

A kind of CP-violating effects in the process $e^+ e^-\to \tau^+ \tau^-$
at Z-pole are parameterized by the $\tau$-lepton weak dipole moment£¬
$d_{w}^{\tau}$£¬ via
the effective vertex for $Z\tau^+ \tau-$ coupling as follows
\begin{equation}
\mathcal{L}_{eff} = ie\bar \tau [ \gamma^\mu V_Z + \gamma^\mu
\gamma_5 A_Z +  \frac{d_{w}^{\tau}}{e} k_\nu \sigma^{\mu \nu} \gamma_5 ]\tau Z_\mu\, ,
\label{Vffcoup}
\end{equation}
where
$$\sigma^{\mu\nu} =\frac{i}{2}\left [\gamma^\mu \gamma^\nu -\gamma^\nu \gamma^\mu \right ], $$
$k$ is the momentum of the gauge boson $Z$;
\begin{equation}
V_Z =   \frac{-1+4 s_w^2}{4s_wc_w}\, , \ \ \
A_Z =-\frac{1}{4s_wc_w}
\end{equation}
are vector and axial vector couplings of gauge boson $Z$ to the
charged leptons in SM, respectively; $s_w=\sin \theta_w$, $c_w=\cos \theta_w$, where $\theta_w$ is
the weak mixing angle. Note that here we highlight that of $\tau$-lepton, so the
coupling of $Z$-boson to electron is assumed as that predicted
by SM at tree level.

Now let us consider the process
\begin{eqnarray}
&& e^-(p_1) +e^+(p_2)\to  \tau^-(p_{\tau^-}) +\tau^+(p_{\tau^+}) \nonumber \\
&&\to a( q^-) +\bar b (q^+)+\nu_\tau (k_1) + \bar \nu_\tau (k_2)
\label{process1}
\end{eqnarray}
at Z-pole, where $a$ and $\bar b$ are mesons originating from $\tau^-$ and $\tau^+$ decays, respectively.
The relevant Feynman diagram for this process is that in FIG. \ref{Fig:process1}.
In eq.(\ref{process1}), $p_1$ and $p_2$ are $4-$momenta of incoming $e^-$ and $e^+$;
$p_{\tau^-}$ and $p_{\tau^+}$ are $4-$momenta of outgoing $\tau^-$ and $\tau^+$ leptons;
$q^-$, $q^+$, $k_1$ and $k_2$ are $4-$momenta of particles $a$, $\bar b$
and neutrinos $\nu_\tau$, $\bar \nu_\tau$ in final state, respectively.
In this paper we also denote $\vec p$ as the $3-$momentum of incoming $e^-$,
$\vec p_{\tau^-}$ as the $3-$momentum of outgoing $\tau^-$,
$\vec q^-$ and $\vec q^+$ as the $3-$momenta of final particles $a$ and $\bar b$ respectively.
Moreover, the reference frame here is chosen such that the incoming $e^-$ is along with the $z-$axis, and
the momentum of the outgoing $\tau^-$ is in the $x-z$ plane, and the angel $\theta_{\tau^-}$ is determined by
the two momenta $\vec{p}_{\tau^-}$ and $\vec{p}$ as $\cos\theta_{\tau^-} = \frac{\vec{p}_{\tau^-}}{|\vec{p}_{\tau^-}|} \cdot
\frac{\vec{p}}{|{\vec p}|}$ (see FIG.\ref{frame} in the Appendix).
The amplitude for the process is expressed as
\begin{eqnarray}
M_1^{a\bar b} &=& ie \bar v (p_2)\gamma^\mu \left ( V_Z + \gamma^5 A_Z\right ) u (p_1)\chi_{\mu\nu}^Z\nonumber \\
&& \frac{G}{\sqrt 2}\bar u (k_1)\gamma^\alpha \left ( 1 - \gamma^5 \right )J_\alpha^a  \chi^{\tau^-}\nonumber  \\
&& ie \left (\gamma^\nu V_Z + \gamma^\nu\gamma_5 A_Z +  \frac{d_{w}^{\tau}}{e} k_\delta \sigma^{\delta \nu} \gamma_5 \right ) \nonumber \\
&& \chi^{\tau^+}\frac{G}{\sqrt 2}\gamma^\beta \left ( 1 - \gamma^5 \right )v (k_2)J_\beta^{\bar b} \, .
\label{matrix1}
\end{eqnarray}
Here $G$ is the Fermi coupling constant, $\chi_{\mu\nu}^Z$ is the propagator
of gauge boson $Z$:
\begin{figure}[htbp]
\begin{tabular}{c}
\includegraphics[width=0.50\linewidth]{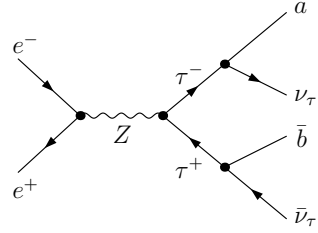}
\end{tabular}
  \caption{Feynman diagram for process $e^- e^+ \to \tau^- \tau^+ \to a \bar b \nu_\tau \bar\nu_\tau$.\label{Fig:process1}}
\end{figure}
\begin{equation}
\chi_{\mu\nu}^Z = \frac{i ( g_{\mu\nu} - \frac{k_\mu k_\nu}{m_Z^2})} {k^2 - m_Z^2+ {\rm i} m_Z \Gamma_Z }\, ,
\end{equation}
$\Gamma_Z$ is the total decay width of gauge boson $Z$ and $k=p_1+p_2$ is the four momentum of $Z$, i.e.
$k^2 = S$ where $\sqrt S$ is the center-of-mass energy. Moreover,
\begin{eqnarray}
\chi^{\tau^-} &=& \frac{i(\not p_{\tau^-} + m_\tau)}{p_{\tau^-}^2 - m_\tau^2 + i m_\tau \Gamma_\tau}\, , \nonumber \\
\chi^{\tau^+} &=& \frac{i(-\not p_{\tau^+} + m_\tau)}{p_{\tau^+}^2 - m_\tau^2 + i m_\tau \Gamma_\tau}\, ,
\end{eqnarray}
where $\Gamma_\tau$ is the total decay width of $\tau$ leptons, are the propagators
of leptons $\tau^-$ and $\tau^+$ respectively. In the expression, the weak hadronic current
$J_\alpha^{\pi^-}$, $J_\beta^{\pi^+}$, $J_\alpha^{\rho^-}$ and $J_\beta^{\rho^+}$ in eq.(\ref{matrix1}) are
as following,
\begin{eqnarray}
J_\alpha^{\pi^-} = f_\pi V_{ud} q_\alpha^-\, ,  \ \ \ &&
J_\beta^{\pi^+} =  f_\pi V_{ud}^* q_\beta^+\, ,\nonumber \\
J_\alpha^{\rho^-} =  f_\rho V_{ud} \varepsilon_\alpha^*\, , \ \ &&
J_\beta^{\rho^+} = f_\rho V_{ud}^* \varepsilon_\beta ,
\label{hadronic-c}
\end{eqnarray}
where $f_\pi$ is the decay constant of $\pi$, $f_\rho$ is the decay constant of $\rho$,
$V_{ud}$ is the CKM matrix element,  $\varepsilon$ is the polarization vector of $\rho$.

Then we shift to consider the pure leptonic decays of $\tau$-lepton pair instead of the semileptonic decays, namely, the process
\begin{eqnarray}
&&e^-(p_1) +e^+(p_2)\to \tau^-(p_{\tau^-}) +\tau^+(p_{\tau^+}) \nonumber \\
&& \to a( q^-) + \bar b (q^+)+ \nu_\tau (k_1) + \bar \nu_\tau (k_2)\nonumber \\
&& +\bar \nu_a (k_3)+ \nu_b (k_4)
\label{process2}
\end{eqnarray}
at Z-pole, where $a$ and $b$ are leptons. The corresponding amplitude
for the process has the same form as that in eq.(\ref{matrix1}),
but with different expressions for $J_\alpha^a$ and $J_\beta^{\bar b}$,
\begin{eqnarray}
J_\alpha^a &=&\bar u (q^-)\gamma_\alpha (1-\gamma^5)v (k_3)\, ,\nonumber \\
J_\beta^{\bar b} &= &\bar u(k_4)\gamma_\beta (1-\gamma^5) v (q^+)\, .
\label{leptonic-c}
\end{eqnarray}
The differential cross section of processes in eq.(\ref{process1}) and eq.(\ref{process2})
is expressed as
\ba
d\sigma_1^{a\bar b}=\frac{1}{2s}\overline {|M_1^{a\bar b}|^2} dLips_{2\to n}.
\label{eq:sigma1}
\ea
$dLips_{2\to n}$ is the phase space of $2\to n $ processes, $n=4$ for process in eq.(\ref{process1}),
$n=6$ for process in eq.(\ref{process2}). $\overline {|M_1^{a\bar b}|^2}$ is the amplitude squared
averaging over spins of initial states and summing over spins of final states.

The observable, relating concerned asymmetry, which involves only the unit momenta of the final particles
$a$ and $\bar b$ may be defined as follows:
\begin{equation}
A_1^{a\bar b} = \frac{\sigma_1^{a\bar b}(+) -\sigma_1^{a\bar b}(-) }{\sigma_1^{a\bar b}(+) +\sigma_1^{a\bar b}(-) },
\label{asy1ab}
\end{equation}
where the cross section $\sigma_1^{a\bar b}(+)$ and $\sigma_1^{a\bar b}(-)$ are defined as
\begin{eqnarray}
\sigma_1^{a\bar b}(+) &=& \sigma_1^{a\bar b}\left ((\hat q^+ \times \hat q^-)\cdot \hat p > 0\right )\, ,\nonumber \\
\sigma_1^{a\bar b}(-) &=& \sigma_1^{a\bar b}\left ((\hat q^+ \times \hat q^-)\cdot \hat p < 0\right )\, . \nonumber
\end{eqnarray}
Here $\hat q^{-} = \frac{\vec {q}^{-}}{|\vec {q}^{-}|}$ is the flying direction of particles $a$,
$\hat q^{+} = \frac{\vec {q}^{+}}{|\vec {q}^{+}|}$ is the flying direction of particles $\bar b$.
$\hat p = \frac{\vec p}{|\vec p|}$ is the direction of the incoming $e^-$ beam.
The observable $A_1^{a\bar b}$ is proportional to the weak dipole moment $d_w^\tau$ when the final charged particles
$a$ and $b$ are the same type of particles, such as $\pi^-$ and $\pi^+$, ($\rho^-$ and $\rho^+$, etc,) in the final state.
Note that SM does also contribute to $A_{1}^{a\bar b}$, when the types of particles $a$ and $b$ are different from
each other, such as $\rho^-$ and $\pi^+$, ($\pi^-$ and $\rho^+$, etc,) in the final states.

It is necessary to consider the CP-conjugated processes for eq.(\ref{process1}) and eq.(\ref{process2}):
$$ e^-(p_1) +e^+(p_2)\to  \tau^-(p_{\tau^-}) +\tau^+(p_{\tau^+}) $$
$$\to b( q^-) +\bar a (q^+)+\nu_\tau (k_1) + \bar \nu_\tau (k_2)$$
and
\begin{eqnarray}
&&e^-(p_1) +e^+(p_2)\to \tau^-(p_{\tau^-}) +\tau^+(p_{\tau^+}) \nonumber \\
&& \to b( q^-) + \bar a (q^+)+ \nu_\tau (k_1) + \bar \nu_\tau (k_2)\nonumber \\
&& +\bar \nu_b (k_3)+ \nu_a (k_4)\, .
\label{process-CJ}
\end{eqnarray}
The amplitude for eq.(\ref{process-CJ}), $M_1^{b\bar a}$, can be obtained by the following replacements,
\begin{equation}
M_1^{b\bar a}=M_1^{a\bar b}\left ({J_\alpha^a \to J_\alpha^b\, ,  J_\beta^{\bar b} \to J_\beta^{\bar a}} \right )\, .
\end{equation}
The corresponding differential cross section $d\sigma_1^{b\bar a}$ is calculated
similarly to $d\sigma_1^{a\bar b}$, see eq.(\ref{eq:sigma1}).
The observable which is analogous to that in eq.(\ref{asy1ab}) is defined as
\begin{equation}
A_1^{b\bar a} = \frac{\sigma_1^{b\bar a}(+) -\sigma_1^{b\bar a}(-) }{\sigma_1^{b\bar a}(+) +\sigma_1^{b\bar a}(-) },
\label{asy1ba}
\end{equation}
where $\sigma_1^{b\bar a}(+)$ and $\sigma_1^{b\bar a}(-)$ are expressed as following
\begin{eqnarray}
\sigma_1^{b\bar a}(+) &=& \sigma_1^{b \bar a}\left ((\hat q^+ \times \hat q^-)\cdot \hat p > 0 \right )\, , \nonumber  \\
\sigma_1^{b\bar a}(-) &=& \sigma_1^{b \bar a}\left ((\hat q^+ \times \hat q^-)\cdot \hat p < 0\right )\, . \nonumber
\end{eqnarray}
It is found that in the sum of $A_{1}^{a\bar b}$ and $A_{1}^{b\bar a}$ the contribution from SM is canceled.
Thus we may define a new general observable as follows:
\begin{equation}
A_1 \equiv \frac{1}{2}\left (A_{1}^{a\bar b} + A_{1}^{b\bar a} \right )\, .
\label{asy11}
\end{equation}
It is proportional to the weak dipole moment $d_w^\tau$ and can be written as a linear function of real and imaginary parts
of $d_w^\tau$:
\begin{equation}
A_{1}= \frac{m_Z}{e}\left [f_1 {\rm Re}(d_w^{\tau}) +g_1 {\rm Im}(d_w^{\tau}) \right ]\, ,
\label{asy12}
\end{equation}
where the slope coefficients $f_1$ and $g_1$ are dimensionless and can be calculated numerically.

Note that these coefficients $f_1$ and $g_1$ for observable $A_1$ are quite small that
it is almost impossible to 'pick up' the signal from backgrounds experimentally. Fortunately,
the observable $A_1$ has a character, that it, approximately, varies as a
trigonometric function $\sin(\cos\theta_{\tau^-})$, so it seems that this property may be applied
to the measurement of the weak dipole moment via constructing the observables so as to increase the
sensitivity (to enhance the signals). Indeed the property can be applied
to enhance the signals and the precision by introducing some new observables,
for which the relevant phase space according to $cos\theta_{\tau^-}>0$ and $cos\theta_{\tau^-}<0$
are divided into two pieces. Such as the cross sections $\sigma_1^{a\bar b}$
for the processes in eq.(\ref{process1}) and eq.(\ref{process2}) are divided
as below:
\begin{eqnarray}
\sigma_1^{a\bar b}(++) &=& \sigma_1^{a\bar b}\left (\cos\theta_{\tau^-}>0, (\hat q^+ \times \hat q^-)\cdot \hat p > 0\right ), \nonumber \\
\sigma_1^{a\bar b}(+-) &=& \sigma_1^{a\bar b}\left (\cos\theta_{\tau^-}>0, (\hat q^+ \times \hat q^-)\cdot \hat p < 0\right ), \nonumber \\
\sigma_1^{a\bar b}(-+) &=& \sigma_1^{a\bar b}\left (\cos\theta_{\tau^-}<0, (\hat q^+ \times \hat q^-)\cdot \hat p > 0\right ), \nonumber \\
\sigma_1^{a\bar b}(--) &=& \sigma_1^{a\bar b}\left (\cos\theta_{\tau^-}<0, (\hat q^+ \times \hat q^-)\cdot \hat p < 0\right ). \nonumber
\label{eq:sigma2}
\end{eqnarray}
We define the observables $A_1^{a\bar b} (+)$ and $A_1^{a\bar b} (-)$:
\begin{eqnarray}
A_1^{a\bar b} (+)&=& \frac{\sigma_1^{a\bar b}(++) -\sigma_1^{a\bar b}(+-) }{\sigma_1^{a\bar b}(++) +\sigma_1^{a\bar b}(+-) }, \nonumber \\
A_1^{a\bar b} (-)&=& \frac{\sigma_1^{a\bar b}(-+) -\sigma_1^{a\bar b}(--) }{\sigma_1^{a\bar b}(-+) +\sigma_1^{a\bar b}(--) }.
\label{asy1ab+}
\end{eqnarray}
In a similar way, the observables $A_1^{b\bar a} (+)$ and $A_1^{b\bar a} (-)$ related to processes in eq.(\ref{process-CJ}) are also defined:
\begin{eqnarray}
A_1^{b\bar a} (+)&=& \frac{\sigma_1^{b\bar a}(++) -\sigma_1^{b\bar a}(+-) }{\sigma_1^{b\bar a}(++) +\sigma_1^{b\bar a}(+-) }, \nonumber \\
A_1^{b\bar a} (-)&=& \frac{\sigma_1^{b\bar a}(-+) -\sigma_1^{b\bar a}(--) }{\sigma_1^{b\bar a}(-+) +\sigma_1^{b\bar a}(--) },
\label{asy1ba+}
\end{eqnarray}
here $\sigma_1^{b\bar a}(++)$, $\sigma_1^{b\bar a}(+-)$, $\sigma_1^{b\bar a}(-+)$ and $\sigma_1^{b\bar a}(--)$ are as following,
\begin{eqnarray}
\sigma_1^{b\bar a}(++) &=& \sigma_1^{b\bar a}\left (\cos\theta_{\tau^-}>0, (\hat q^+ \times \hat q^-)\cdot \hat p > 0\right ), \nonumber \\
\sigma_1^{b\bar a}(+-) &=& \sigma_1^{b\bar a}\left (\cos\theta_{\tau^-}>0, (\hat q^+ \times \hat q^-)\cdot \hat p < 0\right ), \nonumber \\
\sigma_1^{b\bar a}(-+) &=& \sigma_1^{b\bar a}\left (\cos\theta_{\tau^-}<0, (\hat q^+ \times \hat q^-)\cdot \hat p > 0\right ), \nonumber \\
\sigma_1^{b\bar a}(--) &=& \sigma_1^{b\bar a}\left (\cos\theta_{\tau^-}<0, (\hat q^+ \times \hat q^-)\cdot \hat p < 0\right ). \nonumber
\end{eqnarray}
In order to cancel the contribution from SM, we further define new observables as follows:
\begin{eqnarray}
A_1(+) &\equiv & \frac{1}{2}\left (A_{1}^{a\bar b} (+) + A_{1}^{b\bar a} (+) \right ), \nonumber \\
A_1(-) &\equiv & \frac{1}{2}\left (A_{1}^{a\bar b}(-) + A_{1}^{b\bar a} (-)\right ).
\label{asy1ab1}
\end{eqnarray}
Moreover, a new type of observable, which will greatly enhance the signal, is constructed,
\begin{equation}
A_1^\prime \equiv  A_1(+) - A_1 (-).
\label{asy1prime}
\end{equation}
It depends on the real and imaginary parts of the weak dipole moment $d_w^\tau$ linearly:
\begin{equation}
A_1^\prime= \frac{m_Z}{e}\left [f_1^\prime {\rm Re}(d_w^{\tau}) +g_1^\prime {\rm Im}(d_w^{\tau}) \right ]\, ,
\label{asy1primef}
\end{equation}
where the coefficients $f_1^\prime$ and $g_1^\prime$ are dimensionless and can be evaluated by numerical calculations.

Another new observable, which is expected to be more sensitive to the imaginary part of the weak dipole moment,
is defined by a T-even operator $(\hat q^+ + \hat q^-)\cdot \hat p$:
\begin{eqnarray}
A_2^{a\bar b} &=& \frac{\sigma_2^{a\bar b}(+) -\sigma_2^{a\bar b}(-) }{\sigma_2^{a\bar b}(+) +\sigma_2^{a\bar b}(-) }, \nonumber \\
A_2^{b\bar a} &=& \frac{\sigma_2^{b\bar a}(+) -\sigma_2^{b\bar a}(-) }{\sigma_2^{b\bar a}(+) +\sigma_2^{b\bar a}(-) },
\label{asy21}
\end{eqnarray}
where the cross sections $\sigma_2^{a\bar b}(+)$, $\sigma_2^{a\bar b}(-)$, $\sigma_2^{b\bar a}(+)$ and $\sigma_2^{b\bar a}(-)$ are defined as
\begin{eqnarray}
\sigma_2^{a\bar b}(+) &= & \sigma_1^{a\bar b}\left ( (\hat q^+  + \hat q^-)\cdot \hat p > 0\right )\, , \nonumber \\
\sigma_2^{a\bar b}(-) &= & \sigma_1^{a \bar b}\left ((\hat q^+  + \hat q^-)\cdot \hat p < 0\right )\, , \nonumber \\
\sigma_2^{b\bar a}(+) &= & \sigma_1^{b \bar a }\left ((\hat q^+  + \hat q^-)\cdot \hat p > 0 \right )\, ,
\nonumber \\
\sigma_2^{b\bar a}(-) &= & \sigma_1^{b \bar a}\left ((\hat q^+  + \hat q^-)\cdot \hat p < 0\right )\, . \nonumber
\end{eqnarray}
The observable $A_2^{a\bar b}$ has the similar properties as $A_1^{a\bar b}$,
i.e. the contribution from SM is canceled when $a$ and $b$ are same type of particles,
while when the type of particle $a$ is different from that of
particle $b$, SM does contribute some to this observable. Whereas if we add $A_{2}^{a\bar b}$ and
$A_{2}^{b\bar a}$ together, the contribution from SM is canceled. The new observable
is proportional to the weak dipole moment $d_w^\tau$ and can be expressed as
a linear function of its real and imaginary parts:
\begin{eqnarray}
A_{2} &\equiv & \frac{1}{2}\left (A_{2}^{a\bar b}+A_{2}^{b\bar a}\right ) \nonumber \\
&=& \frac{m_Z}{e}\left [f_{2} {\rm Re}(d_w^{\tau}) +g_{2} {\rm Im}(d_w^{\tau}) \right ]\, .
\label{asy22}
\end{eqnarray}
The slope coefficients $f_2$ and $g_2$ are dimensionless and can be evaluated from the theoretical calculation.

\begin{figure}[htbp]
\begin{tabular}{cc}
\includegraphics[width=0.48\linewidth]{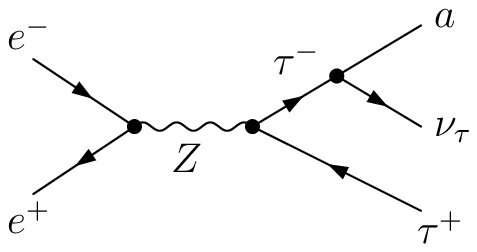} \ & \ \includegraphics[width=0.48\linewidth]{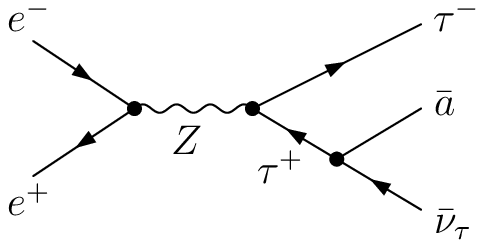}
\end{tabular}
\caption{Feynman diagrams for processes $e^- e^+ \to a \nu_\tau \tau^+$ (left) and $e^- e^+ \to \tau^- \bar a \bar \nu_\tau$ (right).\label{Fig:process3}}
\end{figure}

We also consider processes
$$e^-(p_1) +e^+(p_2)\to  \tau^-(p_{\tau^-}) +\tau^+(p_{\tau^+})$$
$$\to a( q^-) +\nu_\tau (k_1) + \tau^+(p_{\tau^+})$$
and
\begin{eqnarray}
&& e^-(p_1) +e^+(p_2)\to \tau^-(p_{\tau^-}) +\tau^+(p_{\tau^+}) \nonumber \\
&& \to a( q^-) + \nu_\tau (k_1) +\bar \nu_a (k_3)+ \tau^+(p_{\tau^+})
\label{process3}
\end{eqnarray}
at Z-pole. Here the particle $a$ denotes a meson in the first process, while
in the second process it denotes a proper lepton.
The diagrams for the first process in eq.(\ref{process3}) and its CP-conjugated process are put in FIG. \ref{Fig:process3}.
The amplitudes for the processes in eq.(\ref{process3}) can be written as
\begin{eqnarray}
M_3^{a} &=& ie \bar v (p_2)\gamma^\mu \left ( V_Z + \gamma^5 A_Z\right ) u (p_1)\chi_{\mu\nu}^Z\nonumber \\
&& \frac{G}{\sqrt 2}\bar u (k_1)\gamma^\alpha \left ( 1 - \gamma^5 \right )J_\alpha^a  \chi^{\tau^-} ie \left (\gamma^\nu V_Z \right. \nonumber \\
&& \left. +\gamma^\nu\gamma_5 A_Z +  \frac{d_{w}^{\tau}}{e} k_\delta \sigma^{\delta \nu} \gamma_5 \right ) v (p_{\tau^+})\, ,
\label{matrix3}
\end{eqnarray}
and the expression here for the weak hadronic current $J_\alpha^a$ can be found in eq.(\ref{hadronic-c}) or eq.(\ref{leptonic-c}).

The corresponding CP-conjugated processes to eq.(\ref{process3})
$$ e^-(p_1) +e^+(p_2)\to  \tau^-(p_{\tau^-}) +\tau^+(p_{\tau^+})$$
$$ \to \tau^-(p_{\tau^-})  + \bar a ( q^+) +\bar \nu_\tau (k_2) $$
and
\begin{eqnarray}
&& e^-(p_1) +e^+(p_2)\to \tau^-(p_{\tau^-}) +\tau^+(p_{\tau^+}) \nonumber \\
&& \to \tau^-(p_{\tau^-}) + \bar a( q^+) + \bar \nu_\tau (k_2) + \nu_a (k_4)
\label{process3-CJ}
\end{eqnarray}
are also investigated. The relevant amplitude can be expressed:
\begin{eqnarray}
M_3^{\bar a} &=& ie \bar v (p_2)\gamma^\mu \left ( V_Z + \gamma^5 A_Z\right ) u (p_1)\chi_{\mu\nu}^Z \bar u (p_{\tau^-})\nonumber \\
&& ie \left (\gamma^\nu V_Z + \gamma^\nu\gamma_5 A_Z +  \frac{d_{w}^{\tau}}{e} k_\delta \sigma^{\delta \nu} \gamma_5 \right ) \nonumber \\
&& \chi^{\tau^+}\frac{G}{\sqrt 2}\gamma^\beta \left ( 1 - \gamma^5 \right )v (k_2)J_\beta^{\bar a} \, ,
\label{matrix3-CJ}
\end{eqnarray}
where the definition for $J_\beta^{\bar a}$ is that in eq.(\ref{hadronic-c}) or eq.(\ref{leptonic-c}).
The differential cross sections $d\sigma_3^a$ and $d\sigma_3^{\bar a }$ can be calculated in terms of the matrix element $M_3^a$ and
$M_3^{\bar a}$, respectively.

One more new observable, which depends on both the unit momenta of $\tau^-$ and $\tau^+$ and their decay products, can be defined:
\begin{eqnarray}
A_{3}^{a} &=& \frac{\sigma_3^{a}(+) -\sigma_3^{a}(-) }{\sigma_3^{a}(+) +\sigma_3^{a}(-) }, \nonumber \\
A_3^{\bar a} &=& \frac{\sigma_3^{\bar a}(+) -\sigma_3^{\bar a}(-) }{\sigma_3^{\bar a}(+) +\sigma_3^{\bar a}(-) },
\label{asy3}
\end{eqnarray}
\begin{eqnarray}
\sigma_3^a (+) & = & \sigma_3^a((\hat p \times \hat p_{\tau^-})\cdot \hat q^->0), \nonumber \\
\sigma_3^a (-) & = & \sigma_3^a((\hat p \times \hat p_{\tau^-})\cdot \hat q^-<0), \nonumber \\
\sigma_3^{\bar a} (+) & = & \sigma_3^{\bar a}((\hat p \times \hat p_{\tau^-})\cdot \hat q^+>0), \nonumber \\
\sigma_3^{\bar a} (-) & = & \sigma_3^{\bar a}((\hat p \times \hat p_{\tau^-})\cdot \hat q^+<0).
\end{eqnarray}
Here $\hat p_{\tau^-} = \frac{\vec p_{\tau^-}}{|\vec p_{\tau^-}|}$ is the unit momentum of $\tau^-$,
Note that SM does contribute some to the observable $A_3^{a}$ and $A_3^{\bar a}$,
but according to the CP property of processes in eq.(\ref{process3}) and eq.(\ref{process3-CJ}),
one may conclude that the contribution from SM in $A_{3}^{a} +  A_{3}^{\bar a}$ is canceled, and
similar discussions can be found in ref. \citep{Yang:1997iv}.
Thus the new general observable can be defined and again it depends on the real and imaginary
parts of weak dipole moment $d_w^\tau$ linearly:
\begin{eqnarray}
A_3 &\equiv & \frac{1}{2}\left (A_{3}^{a} + A_{3}^{\bar a} \right ) \nonumber \\
&=&\frac{m_Z}{e}\left [f_3 {\rm Re}(d_w^{\tau}) +g_3 {\rm Im}(d_w^{\tau}) \right ]\, .
\label{asy31}
\end{eqnarray}
Here $f_3$ and $g_3$ are dimensionless coefficients, which can be determined numerically from theoretical
calculations.

Note that these slope coefficients $f_3, g_3$ for observable $A_3$ are so small that
it is nearly impossible to make difference between the signal and the background,
similar to the case of observable $A_1$. Following the technique dealing with $A_1$,
we separate the phase space into two parts: $cos\theta_{\tau^-}>0$ and $cos\theta_{\tau^-}<0$.
To investigate the observable $A_3^a$ by dividing the phase space,
the total cross sections $\sigma_3^a$ for processes in eq.(\ref{process3})
are then divided into four parts,
\begin{eqnarray}
\sigma_3^a(++)=\sigma_3^a\left (\cos\theta_{\tau^-}>0,(\hat p \times \hat p_{\tau^-})\cdot \hat q^- > 0\right ), \nonumber \\
\sigma_3^a(+-)=\sigma_3^a\left (\cos\theta_{\tau^-}>0,(\hat p \times \hat p_{\tau^-})\cdot \hat q^- < 0\right ), \nonumber \\
\sigma_3^a(-+)=\sigma_3^a\left (\cos\theta_{\tau^-}<0,(\hat p \times \hat p_{\tau^-})\cdot \hat q^- > 0\right ), \nonumber \\
\sigma_3^a(--)=\sigma_3^a\left (\cos\theta_{\tau^-}<0,(\hat p \times \hat p_{\tau^-})\cdot \hat q^- < 0\right ). \nonumber
\end{eqnarray}
Observables $A_3^a(+)$ and $A_3^a (-)$ are defined in terms of the division $cos\theta_{\tau^-}>0$ and $cos\theta_{\tau^-}<0$, respectively
\begin{eqnarray}
A_3^a (+)&=& \frac{\sigma_3^a(++) -\sigma_3^a(+-) }{\sigma_3^a(++) +\sigma_3^a(+-) }, \nonumber \\
A_3^a (-)&=& \frac{\sigma_3^a(-+) -\sigma_3^a(--) }{\sigma_3^a(-+) +\sigma_3^a(--) }.
\label{asy3a+}
\end{eqnarray}
Similarly observables $A_3^{\bar a} (+)$ and $A_3^{\bar a} (-)$ relating to the processes in eq.(\ref{process3-CJ}) are defined
\begin{eqnarray}
A_3^{\bar a} (+)&=& \frac{\sigma_3^{\bar a}(++) -\sigma_3^{\bar a}(+-) }{\sigma_3^{\bar a}(++) +\sigma_3^{\bar a}(+-) }, \nonumber \\
A_3^{\bar a} (-)&=& \frac{\sigma_3^{\bar a}(-+) -\sigma_3^{\bar a}(--) }{\sigma_3^{\bar a}(-+) +\sigma_3^{\bar a}(--) },
\label{asy3abar+}
\end{eqnarray}
$\sigma_3^{\bar a}(++)$, $\sigma_3^{\bar a}(+-)$, $\sigma_3^{\bar a}(-+)$ and $\sigma_3^{\bar a}(--)$ are listed in the following,
\begin{eqnarray}
\sigma_3^{\bar a}(++)=\sigma_3^{\bar a}\left (\cos\theta_{\tau^-}>0,(\hat p \times \hat p_{\tau^-})\cdot \hat q^+ > 0\right ), \nonumber \\
\sigma_3^{\bar a}(+-)=\sigma_3^{\bar a}\left (\cos\theta_{\tau^-}>0,(\hat p \times \hat p_{\tau^-})\cdot \hat q^+ < 0\right ), \nonumber \\
\sigma_3^{\bar a}(-+)=\sigma_3^{\bar a}\left (\cos\theta_{\tau^-}<0,(\hat p \times \hat p_{\tau^-})\cdot \hat q^+ > 0\right ), \nonumber \\
\sigma_3^{\bar a}(--)=\sigma_3^{\bar a}\left (\cos\theta_{\tau^-}<0,(\hat p \times \hat p_{\tau^-})\cdot \hat q^+ < 0\right ). \nonumber
\end{eqnarray}
To cancel the contribution from SM, the general observables are defined as following,
\begin{eqnarray}
A_3(+) &\equiv & \frac{1}{2}\left (A_{3}^{a} (+) + A_{3}^{\bar a} (+) \right ), \nonumber \\
A_3(-) &\equiv & \frac{1}{2}\left (A_{3}^{a}(-) + A_{3}^{\bar a} (-)\right ).
\label{asy3abara}
\end{eqnarray}
A new type of observable which will greatly enhance the signal are constructed,
\begin{equation}
A_3^\prime  \equiv A_3(+) - A_3(-),
\label{asy3prime}
\end{equation}
which can be expressed as a linear function as the real and imaginary parts of the weak dipole moment $d_w^\tau$,
\begin{equation}
A_3^\prime= \frac{m_Z}{e}\left [f_3^\prime {\rm Re}(d_w^{\tau}) +g_3^\prime {\rm Im}(d_w^{\tau}) \right ]\, ,
\label{asy3primef}
\end{equation}
the dimensionless coefficients $f_3^\prime$ and $g_3^\prime$ can be determined numerically.

In this section three types of new observables $A_1^\prime$, $A_2$ and $A_3^\prime$ are constructed via operators $(\hat q^+ \times \hat q^-)\cdot \hat p$,
$(\hat q^+ + \hat q^-)\cdot \hat p$ and $(\hat p \times \hat p_{\tau^-})\cdot \hat q^\pm$, respectively.
These new observables are found to be proportional to the real and imaginary parts of the weak dipole moment $d_w^\tau$,
with slope coefficients dependent on the SM coupling of gauge boson $Z$ to leptons, the kinematics of processes and the decay modes of $\tau$ leptons.
These coefficients can be calculated numerically for different decay products of $\tau^-$ and $\tau^+$.
Once these new-constructed observables are measured, the real and imaginary parts of the weak dipole moment $d_w^\tau$ can be extracted
by solving these equations in eqs.(\ref{asy1primef},\ref{asy22},\ref{asy3primef}).

\section{The observables employed by LEP-I for measuring $d_w^\tau$}
To determine the weak dipole moment $d_w^\tau$ at $Z-$pole, several observables and methods
were employed at LEP-I.
In this section some of the observables used at LEP-I are reviewed briefly.

In refs.\citep{Bernreuther:1989kc,Bernreuther:1991xe,Bernreuther:1993nd},
the CP-odd tensor observables, which are constructed by the momenta or unit momenta of the decay
products of $\tau^+$ and $\tau^-$:
$$T_{ij} = \left ( \vec q^+ - \vec q^-\right )_{i} \left ( \vec q^+ \times \vec q^-\right )_j + \left ( i \to j \right )$$
or
\begin{equation}
\hat T _{ij}= \left ( \hat q^+ - \hat q^-\right )_{i} \frac
{\left ( \hat q^+ \times \hat q^-\right )_j}{|\hat q^+ \times \hat q^-|}+ \left ( i \to j \right )
\label{operator1}
\end{equation}
are studied carefully. Here $i,j = 1, 2, 3$ are Cartesian vector indices.
Expectation values for them are proportional to the real part
of the weak dipole moment $d_w^\tau$ as
\begin{eqnarray}
\langle T_{ij}\rangle& =& \frac{m_Z}{e}c s_{ij}{\rm Re}(d_w^{\tau})\,, \nonumber \\
\langle\hat T _{ij}\rangle & = & \frac{m_Z}{e}\hat c s_{ij}{\rm Re}(d_w^{\tau})\,.
\label{operator1-coef}
\end{eqnarray}
Here $s_{ij}$ is the tensor polarization of the intermediate $Z$ state,
which can be written as diag$\left (-\frac{1}{6}, -\frac{1}{6}, \frac{1}{3} \right )$ if
the beam direction is identified with the three-axis.
The coefficients $c$ and $\hat c$ are calculated precisely in ref. \citep{Bernreuther:1993nd}.
Note that we have checked the results i.e. the same values for these coefficients as those in ref. \citep{Bernreuther:1993nd} are obtained. By measuring observables in eq.(\ref{operator1}),
an up bound for the real part of the weak dipole moment $d_w^\tau$ was obtained by OPAL and
ALEPH collaboration at LEP-I \citep{Buskulic:1992jj, Acton:1992ff, Buskulic:1994yh}.
Note that the advantage of this method is
that only the flying directions of $\tau$ decay products are enough for the measurement,
namely, any information about the momenta and spins of $\tau$ leptons is not needed.

In ref.\citep{Akers:1994xf}, the so-called optimal T-even and T-odd observables, $\hat Q^+$ and $\hat Q^-$,
are employed to measure the CP-violating effects in process $Z\to \tau^- \tau^+$ by OPAL collaboration.
They are defined as following,
\begin{equation}
\hat Q^+ = \frac{M_{\rm CP}^{\rm Im}}{M_{\rm SM}}, \ \ \
\hat Q^- = \frac{M_{\rm CP}^{\rm Re}}{M_{\rm SM}},
\label{operator2}
\end{equation}
where
\begin{eqnarray}
M_{\rm CP}^{\rm Im} & = & \left (\hat p_{\tau^-}\cdot \hat p \right )\left [ (\hat p_{\tau^-}\cdot \hat s_+)(\hat p \cdot \hat s_-)\right. \nonumber \\
&& \left. -(\hat p_{\tau^-}\cdot \hat s_-)(\hat p \cdot \hat s_+) \right ]\, , \nonumber \\
M_{\rm CP}^{\rm Re} & = &\left (\hat p_{\tau^-}\cdot \hat p \right ) \left ( \hat p_{\tau^-}\times (\hat s_+ - \hat s_-) \right )\cdot \hat p\, , \nonumber \\
M_{\rm SM} & = & 1 + \left (\hat p_{\tau^-}\cdot \hat p \right )^2 + \hat s_+ \cdot \hat s_-\left ( 1 - \left (\hat p_{\tau^-}\cdot \hat p \right )^2 \right )\nonumber \\
&& - 2 (\hat p \cdot \hat s_+)(\hat p \cdot \hat s_-) + 2 \left (\hat p_{\tau^-}\cdot \hat p \right )\nonumber \\
&& \left [ (\hat p_{\tau^-}\cdot \hat s_+)(\hat p \cdot \hat s_-)+(\hat p_{\tau^-}\cdot \hat s_-)(\hat p \cdot \hat s_+) \right ]. \nonumber
\end{eqnarray}
Here $\hat s_+$ and $\hat s_-$ are spin vectors of $\tau^+$ and $\tau^-$ leptons in their respective rest frame, respectively.
\footnote{ Note that the definition of T-even and T-odd operators in section 2 of ref.\citep{Akers:1994xf} are exchanged
in comparison with the definition in eq.(\ref{operator2}),
We suspect that there are misprints in ref.\citep{Akers:1994xf}.}
Note that only the unit momenta and spins of $\tau^-$ and $\tau^+$ leptons
are required in order to obtain the average values for operators in eq.(\ref{operator2}).
Theoretically the expectation values for these T-even and T-odd operators are proportional
to the imaginary part and real part
of $d_w^\tau$, respectively.
\begin{eqnarray}
\langle \hat Q^{+}\rangle & =& \frac{m_Z}{e}g_{\rm LEP}{\rm Im}(d_w^{\tau}), \nonumber \\
\langle\hat Q^{-}\rangle& = & \frac{m_Z}{e}f_{\rm LEP}{\rm Re}(d_w^{\tau}).
\end{eqnarray}
The values of the coefficients $f_{\rm LEP}$ and $g_{\rm LEP}$ corresponding to the different
decay modes of $\tau^-$ and $\tau^+$ leptons were obtained by simulation and presented in
Table 4 of ref.\citep{Akers:1994xf}.
Note that the advantage of the chosen observables is that both of the real and
imaginary parts of the weak dipole moment $d_w^\tau$ can be determined, while only
real part of $d_w^\tau$ is measured by employing operators in eq.(\ref{operator1}).
Moreover, choices of the observables in eq.(\ref{operator2}) can optimize the signal-to-background ratio
in the measurements \citep{Akers:1994xf}.

In ref.\citep{Bernabeu:1993er} the following observables relating to $\tau$ decay products were defined
\begin{eqnarray}
A_{sc}^a & = & \frac{\sigma_{sc}^a (+) -\sigma_{sc}^a (-) }{\sigma_{sc}^a (+) +\sigma_{sc}^a (-) }\, , \nonumber \\
A_{sc}^{\bar a} & = & \frac{\sigma_{sc}^{\bar a} (+) -\sigma_{sc}^{\bar a} (-) }{\sigma_{sc}^{\bar a} (+) +\sigma_{sc}^{\bar a} (-) }\, ,
\end{eqnarray}
where the cross section $\sigma_{sc}^a (+)$, $\sigma_{sc}^a (-)$, $\sigma_{sc}^{\bar a} (+)$ and $\sigma_{sc}^{\bar a} (-)$ expressed as
\begin{eqnarray}
&&\sigma_{sc}^a (+) =\left [ \int_0^1 d(\cos\theta_{\tau^-}) \int_0^\pi d\Phi_a + \right. \nonumber \\
&& \left. \int_{-1}^0 d(\cos\theta_{\tau^-}) \int_\pi^{2\pi} d\Phi_a\right ]\frac{d\sigma_3^a}{d(\cos\theta_{\tau^-})d\Phi_a},\nonumber \\
&&\sigma_{sc}^a (-) =\left [ \int_0^1 d(\cos\theta_{\tau^-}) \int_\pi^{2\pi} d\Phi_a + \right. \nonumber \\
&& \left. \int_{-1}^0 d(\cos\theta_{\tau^-}) \int_0^{\pi} d\Phi_a\right ]\frac{d\sigma_3^a}{d(\cos\theta_{\tau^-})d\Phi_a},\nonumber
\end{eqnarray}
\begin{eqnarray}
&&\sigma_{sc}^{\bar a} (+) =\left [ \int_0^1 d(\cos\theta_{\tau^-}) \int_0^\pi d\Phi_{\bar a} + \right. \nonumber \\
&& \left. \int_{-1}^0 d(\cos\theta_{\tau^-}) \int_\pi^{2\pi} d\Phi_{\bar a}\right ]\frac{d\sigma_3^{\bar a}}{d(\cos\theta_{\tau^-})d\Phi_{\bar a}},\nonumber \\
&&\sigma_{sc}^{\bar a} (-) =\left [ \int_0^1 d(\cos\theta_{\tau^-}) \int_\pi^{2\pi} d\Phi_{\bar a} + \right. \nonumber \\
&& \left. \int_{-1}^0 d(\cos\theta_{\tau^-}) \int_0^{\pi} d\Phi_{\bar a}\right ]\frac{d\sigma_3^{\bar a}}{d(\cos\theta_{\tau^-})d\Phi_{\bar a}}.
\end{eqnarray}
Here $\Phi_a$ and $\Phi_{\bar a}$ are defined in FIG.\ref{frame} in the appendix,
$d\sigma_3^a$ and $d\sigma_3^{\bar a}$ are differential cross sections calculated from the processes in eq. (\ref{process3}) and eq.(\ref{process3-CJ}),
respectively. Both the observables $A_{sc}^a$ and $A_{sc}^{\bar a}$ are free from the backgrounds due to SM and satisfy the relation\citep{Bernabeu:1993er}
$$
A_{sc}^a = A_{sc}^{\bar a} \propto {\rm Re} (d_w^{\tau}).
$$
A genuine CP-violating observable is defined
\begin{equation}
A_{sc}^{\rm CP} = \frac{1}{2}\left ( A_{sc}^a +A_{sc}^{\bar a} \right )\, ,
\label{asysc}
\end{equation}
which can be expressed as the linear function of the real part of the weak dipole moment $d_w^\tau$,
\begin{equation}
A_{sc}^{\rm CP} = \frac{m_Z}{e} f_{sc} {\rm Re} (d_w^{\tau}).
\end{equation}
Here $f_{sc}$ is the dimensionless coefficients which depends on different decay modes of $\tau$ leptons and can be calculated numerically.
Note that the OPAL and L3 collaboration did employ this observable to measure the weak dipole moment of the $\tau$ lepton at LEP-I \citep{Ackerstaff:1996gy, Acciarri:1998zc}.

The latest bound on the real part and imaginary part of $d_{w}^{\tau}$ were determined by the ALPEH collaboration in 2003 \citep{Heister:2002ik}.
This measurement is based on the method of maximum likelihood fit to the date, taking into account
the differential cross section with different spins of $\tau$ leptons, including spin correlation.
The differential cross section of $e^- e^+ \to \tau^- \tau^+$ can be expressed as \citep{Stiegler:1993hi, Heister:2002ik}
\begin{eqnarray}
&&\frac{d\sigma}{d\cos\theta_{\tau^-}}(\hat s_-, \hat s_+) =R_{00} + \sum_{\mu = 1,3} R_{\mu 0}s_-^\mu   \nonumber \\
&& + \sum_{\nu = 1,3} R_{0 \nu}s_+^\nu  + \sum_{\mu, \nu = 1,3} R_{\mu \nu}s_-^\mu s_+^\nu.
\end{eqnarray}
Here $R_{\mu\nu}$ terms are functions of fermion couplings to gauge boson $Z$ and the $\theta_{\tau^-}$.
The contribution from SM is separated from those from non-SM by defining
$$
\left (R_{\mu\nu}\right )_\pm =  R_{\mu\nu} \pm R_{\nu\mu}\, .
$$
Note that both the real and imaginary parts of weak dipole moment $d_w^\tau$ can be determined by
measuring $\left (R_{\mu\nu}\right )_\pm$.

\section{Numerical results and discussions}

In this paper we highlight the measurements of $d_w^\tau$, the weak dipole moment of $\tau$-lepton,
at an $e^+e^-$ collider which runs at $Z$-boson pole (a Z-factory). In contrary to low energy colliders
such as B-factory and BEPC, such a high energy collider at $\sqrt S\simeq m_Z$ has an additional
advantage besides the $Z$-boson resonance effect, for measuring the weak dipole moment.
Namely at such a high energy collider, the
produced $\tau$ and $\bar{\tau}$ gain a great momentum, i.e. the great Lorentz boost
($\gamma \simeq 25.66$), correspondingly, statistically the produced
$\tau$ ($\bar{\tau}$) leptons may travel a few $mm$ before they decay \cite{Heister:2001uh}, so that
the flying directions of the produced $\tau$ and $\bar{\tau}$ leptons at a Z-factory can be reconstructed quite well with
a fine vertex detector by precisely measuring the charged track between the colliding point of $e^+e^-$ and
the decay vertices of the $\tau$ and $\bar{\tau}$ leptons respectively. The so-called 3-dimensional method developed in ref.\cite{Kuhn:1993ra}
is based on the determination of flying directions of the produced $\tau$ and $\bar{\tau}$ leptons and it was widely
employed by experimental groups at LEP-I, for example, the
measurements of the weak dipole moment of $\tau$ lepton \cite{Akers:1994xf, Ackerstaff:1996gy, Acciarri:1998zc, Heister:2002ik},
as well as the lifetime and polarization of the produced $\tau$ lepton \cite{Barate:1997hw,Heister:2001uh}.
Considering the advantage further that the flying directions of the produced $\tau$ and $\bar{\tau}$ leptons
may be constructed at a Z-factory, here we propose new observables as in eq.(\ref{asy1prime}),
eq.(\ref{asy22}) and eq.(\ref{asy3prime}), and two of them, i.e.
observables defined in eq.(\ref{asy1prime}) and eq.(\ref{asy3prime}),
relate to the unit momentum of the produced $\tau$-lepton deeply.

The amplitudes for the production and decays of $\tau$ lepton pair at $e^-e^+$ colliders at $Z-$pole
are calculated using the package FeynArts \citep{Hahn:2000kx} and FormCalc \citep{Hahn:1998yk, Hahn:2004rf}.
New-constructed observables defined in eq.(\ref{asy1prime}), eq.(\ref{asy22}) and eq.(\ref{asy3prime}) are
calculated and their dimensionless coefficients $f_{1}^\prime$ and $g_{1}^\prime $, $f_2$ and $g_2$,
as well as $f_3^\prime$ and $g_3^\prime$ are listed in TABLE \ref{table:coffe-re}.
In order to compare the sensitivity of the observables for measuring
the weak dipole moment $d_w^\tau$, coefficients of those
observables employed at LEP-I are also listed in TABLE \ref{table:coffe-re}.

Firstly, let us consider the observable $\hat T_{33}$ defined in eq.(\ref{operator1}).
Here the values of coefficients $\hat c$ for the decay modes of $\tau$ lepton are
taken from TABLE VI of ref. \citep{Bernreuther:1993nd}.
From eq.(\ref{operator1-coef}) one can see clearly that $\hat c s_{33}$ describes the
sensitivity of the expectation values for the operator $\hat T_{33}$ to
the weak dipole moment $d_w^\tau$. Hence we define
$$\hat c_{33} \equiv \hat c s_{33} = \frac{1}{3}\hat c,$$
and list their values for the concerned decay modes of $\tau$ lepton in TABLE \ref{table:coffe-re}.
Obviously these numerical relation holds
$$|f_1^\prime|>|f_3^\prime|>|\hat c_{33}|.$$
Note that the first term and the second term in eq. (\ref{operator1}) are exactly the same for
$i=j=3$, and hence there exists a constant of $2$ in the definition of $\hat T_{ij}$.
From TABLE \ref{table:coffe-re} it is concluded that our new-constructed observables $A_1^\prime$
and $A_3^\prime$ are much more sensitive to the real part of $d_w^\tau$ than the observable defined by eq. (\ref{operator1}).
\begin{table}[htbp]
\begin{center}
\begin{tabular}{|c|c|c|c|c|}\hline
& $\pi^-\pi^+$ & $\rho^-\rho^+$ & $\pi^-\rho^+$ & $e^-e^+$ \\ \hline
$f_1^\prime$ &$0.938$ &$0.464$ & $0.789$& $-0.322$\\ \hline
$g_1^\prime$ &$-0.0039$ & $-0.0016$& $-0.0073$& $0.0068$\\ \hline
$f_2$ & $0.020$& $0.002$& $0.006$&$-0.002$\\ \hline
$g_2$ & $0.2394$& $0.1129$& $0.2000$& $-0.0545$\\ \hline
$f_3^\prime$ &$-0.815$ & $-0.366$& $\times$ & $0.272$ \\ \hline
$g_3^\prime$ &$0.0048$ & $0.0023$& $\times$& $-0.0016$\\ \hline
$\hat c_{33}$ & $-0.667$ & $-0.304$ & $-0.513$ & $0.209$ \\ \hline
$f_{\rm sc}$ & $-0.408$& $-0.183$& $\times$&$0.136$ \\ \hline
$f_{\rm LEP}$ & $0.201$& $0.211$& $0.204$& $-0.056$\\ \hline
$g_{\rm LEP}$ & $-0.0450$& $-0.0490$& $-0.0265$& $-0.0046$\\ \hline
\end{tabular}
\caption{Values for slope coefficients $f_1^\prime$, $g_1^\prime$, $f_2$, $g_2$, $f_3^\prime$, $g_3^\prime$, $\hat c_{33}$, $f_{\rm sc}$, $f_{\rm LEP}$ and $g_{\rm LEP}$. Here '$\times$' means there is no
definition for the corresponding observables.}
\label{table:coffe-re}
\end{center}
\end{table}

Secondly, let us highlight the observable $A_{\rm sc}^{CP}$ defined in eq.(\ref{asysc}),
which was employed by OPAL and L3 collaborations at LEP-I \citep{Ackerstaff:1996gy, Acciarri:1998zc}.
It is proportional to the real part of weak dipole moment $d_w^\tau$ and
the corresponding coefficient $f_{\rm sc}$ is calculated and listed in TABLE \ref{table:coffe-re}.
Furthermore, the so-called optimal T-even and T-odd operators in eq.(\ref{operator2})
which are used to measure the CP-violating effects in process $Z\to \tau^- \tau^+$
by the OPAL collaboration \citep{Akers:1994xf}, are also concerned here.
Their expectation values are proportional to the real part and imaginary part
of $d_w^\tau$, respectively.  Values for their coefficients $f_{\rm LEP}$ and $g_{\rm LEP}$ are taken
from Table 4 of ref.\citep{Akers:1994xf} and listed in TABLE \ref{table:coffe-re}, too.

From TABLE \ref{table:coffe-re}, one can draw conclusions as follows.
Generally the new-constructed observables depend on both the real and imaginary parts of
the weak dipole moment $d_w^\tau$ linearly.
For the observables defined by T-odd operators, such as $A_1^\prime$ and $A_3^\prime$,
their slope coefficients satisfy these relations
$$|f_1^\prime| \gg |g_1^\prime|,\ \ \  |f_3^\prime| \gg |g_3^\prime|$$
and
$$|f_1^\prime| > |f_3^\prime| > |f_{\rm sc}| > |f_{\rm LEP}|. $$
Especially for $\pi^-,\pi^+$ final states, we have
$$|f_1^\prime| \gtrsim |f_3^\prime| \simeq 2|f_{\rm sc}| \simeq 4|f_{\rm LEP}|.$$
It means that the observables $A_1^\prime$ and $A_3^\prime$ are
much more sensitive to the CP-violated weak dipole moment than the observables
employed at LEP-I. Though these two observables are constructed further by dividing the phase
space relating to the direction of the produced $\tau$ lepton into two pieces, their superiority
will not be reduced, especially, at future colliders with high luminosity
and equipping advanced detectors with very fine vertex detector so as to resolute
the direction of the produced $\tau$ lepton. In principle, when two of the new observables are measured,
then with the coefficients listed in TABLE \ref{table:coffe-re} the real part and imaginary part of
the weak dipole $d_w^\tau$ may be determined separately.

The relations in the following hold for observable $A_2$ which is defined via the T-even operator,
$$|f_2| \ll |g_2|\, , \ \ |g_2| > |g_{\rm LEP}|$$
(see TABLE \ref{table:coffe-re}). For $\pi^-,\pi^+$ final states, we have
$$|g_2| \simeq 5|g_{\rm LEP}|.$$
This implies that the observable $A_2$ depends on the imaginary part of $d_w^\tau$
much more strongly than the real part. Moreover, it is superior to the observable employed
by LEP-I for measurement the imaginary part of the weak dipole moment $d_w^\tau$.

In this paper, the new observables, constructed in terms of CP(T)-odd operators with
$\tau$-lepton phase space being divided, are suggested
for measuring the weak dipole moment $d_w^\tau$ at an $e^+e^-$ collider running at Z-pole.
Then by precisely numerical calculations, it is shown that the new observables are more sensitive
in measuring $d_w^\tau$ than those applied by LEP-I.
They are proportional to the real and imaginary parts of $d_w^\tau$ and the slope coefficients
are much larger than those employed at LEP-I. It implies that the effects caused by the weak dipole
moment $d_w^\tau$ are enhanced greatly once these new observables are employed,
i.e. the signals for the weak dipole moment $d_w^\tau$ are amplified at a $Z$-factory in certain senses
in comparison with the previous observables.

Therefore it is expected that once these new constructed observables are employed in
future measurements or re-analyzing the LEP-I data,
an improved up-bound for the real and imaginary parts of the weak dipole moment $d_w^\tau$
will be reached.
\section*{Acknowledgments}
The author (Q. Xu) would like to thank Prof. Zong-Guo Si for helpful discussions.
This work is supported by the National Natural Science Foundation of China
under Grants Nos. 11275243, 11147001, 11147023, 11305044; the open project of
State Key Laboratory of Theoretical Physics with Grant No. Y3KF311CJ1;
and Zhejiang Provincial Natural Science Foundation of China under Grant No. LQ12A05003.
\section*{Appendix}
\vspace*{-5mm}
\begin{figure}[H]
\begin{center}
\includegraphics[width=0.8\linewidth]{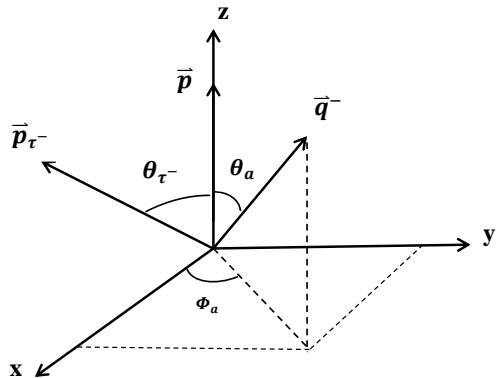}
\caption{The reference frame employed in this paper.}
\label{frame}
\end{center}
\end{figure}
\bibliographystyle{unsrtnat}

\end{document}